\documentclass[twocolumn,showpacs,amsfonts,aps,prc,nofootinbib,floatfix,%
superscriptaddress]{revtex4}

\usepackage{amsmath}
\usepackage{bm}
\usepackage{graphicx}

\usepackage{epsfig}
\newcommand{\beq}{\begin{equation}}
\newcommand{\eeq}{\end{equation}}
\newcommand{\bea}{\vspace{0.25cm}\begin{eqnarray}}
\newcommand{\eea}{\end{eqnarray}}

\newcommand{\ro}{\mbox{{\boldmath
$\rho$}}}

\newcommand{\pb}{{{\bf p}}}

\newcommand{\bb}{{{\bf b}}}

\def\lsim{\mathrel{\rlap{\lower4pt\hbox{\hskip1pt$\sim$}}
    \raise1pt\hbox{$<$}}}         %less than or approx. symbol
\def\gsim{\mathrel{\rlap{\lower4pt\hbox{\hskip1pt$\sim$}}
    \raise1pt\hbox{$>$}}}         %greater than or approx. symbol

\newcommand{\landau}{L.D.~Landau Institute for Theoretical Physics,
        GSP-1, 117940, Kosygina Str. 2, 117334 Moscow, Russia}

\begin{document}

%%%%%%%%%%%%%%%%%%%%%%%%Front Matter%%%%%%%%%%%%%%%%%%%%%%%%%%%%%%%%%%
%%%%%%%%%%%%%%%%%%%%%%%%%%%%%%%%%%%%%%%%%%%%%%%%%%%%%%%%%%%%%%%%%%%%%%
%\renewcommand{\thefootnote}{\fnsymbol{footnote}}

\title{
Flavor dependence of jet quenching in $pp$ collisions
and its effect on the $R_{AA}$ for heavy mesons
}
\date{\today}

\author{B.G.~Zakharov}\affiliation{\landau}

\begin{abstract}
We study the flavor dependence of the medium modification
factor $R_{pp}$ for $pp$ collisions for scenario with formation of a small-size quark-gluon
plasma (QGP) for RHIC ($\sqrt{s}=0.2$ TeV) 
and LHC ($\sqrt{s}=2.76$ TeV) energies. 
We find that at $p_T\sim 10$ GeV
the pion spectrum  is suppressed by $\sim 20-30\,(25-35)$\% for RHIC (LHC), 
for $D$~($B$) mesons the suppression effect is smaller by a factor of $\sim 0.7-0.8$ 
($0.5$).
The flavor hierarchy 
$R_{pp}^\pi<R_{pp}^D<R_{pp}^B$
is held at $p_T\lsim 20$ GeV for RHIC and at $p_T\lsim 70$ GeV
for LHC. 
This gives a sizeable reduction of the 
heavy-to-light ratios of the nuclear modification factors $R_{AA}$
as compared to that in the standard scenario without
the QGP production in $pp$ collisions. 
\end{abstract}
%
%\pacs{12.38.Mh, 24.85.+p}

\maketitle

%%%%%%%%%%%%%%%%%%%OUR%%%%%%%%%%%%%%%%%%%%%%%%%%%%%%%%%%%%%%%%%%%%%%%
\section{Introduction}
There is general consensus that the observed in experiments on
$AA$ collisions at RHIC and LHC suppression of 
the high-$p_{T}$ particle spectra (jet quenching)
is due to radiative and collisional parton energy loss in the QGP
formed in the initial stage of $AA$ collisions.
In the scenario with formation of a mini fireball of the 
QGP in $pp$ collisions a similar mechanism should suppress 
particle spectra as compared to the predictions of the standard
pQCD that neglect the final state interaction (FSI) effects for fast
partons produced in hard reactions. 
In this case the measured $pp$ inclusive
cross section 
will differ from that predicted by the pQCD, 
$d\sigma_{pert}(pp\to hX)/d\pb_{T}dy$, 
by the medium modification
factor $R_{pp}$ 
\beq
R_{pp}=\frac{d\sigma_{}(pp\to hX)/d\pb_{T}dy}
{d\sigma_{pert}(pp\to hX)/d\pb_{T}dy}\,,
\label{eq:10}
\eeq
where the numerator is the real $pp$ inclusive cross section
accounting for the FSI effects in the mini-QGP.
Of course the $R_{pp}$ is not directly measurable quantity,
because the spectrum with the FSI effects switched off
is unknown. 
Nevertheless, it is clearly of great interest for understanding the 
uncertainties of the standard pQCD
predictions due to the FSI higher twist effects. Also, the medium suppression
in $pp$ collisions is important for theoretical predictions for
the nuclear modification factor $R_{AA}$ for $AA$ collisions, because
the $pp$ baseline spectrum, that is necessary for evaluating the $R_{AA}$, 
should account for the medium effects in the mini-QGP.
In principle, the medium modification of jets in $pp$ collisions can be
studied experimentally by measuring the multiplicity dependence of the 
direct photon-triggered fragmentation functions (FFs) \cite{Z_pp}.

The idea that the QGP may be 
produced in $pp$ collisions 
has attracted much attention in recent years
(see, for instance,
\cite{Bozek_pp,Wied_pp,Camp1,Gyulassy_pp,glasma_pp,SZ}). It is
mostly due to observation of the ridge effect in
high multiplicity $pp$ events at
$\sqrt{s}=7$ TeV by the CMS collaboration  \cite{CMS_ridge1},
which may be caused by  the transverse flow of the produced mini-QGP
fireball.
It is important that the conditions for the mini-QGP formation 
in $pp$ collisions are better
in jet events, because the multiplicity of soft off-jet particles 
(the so-called underlying events (UE))
is enhanced by a factor of $\sim 2.5$ \cite{Field}
as compared to minimum bias events.
Therefore, even at RHIC energies $\sqrt{s}\sim 0.2$ TeV 
the UE multiplicity may be high enough for the QGP formation.
The scenario with mini-QGP in jet $pp$ events
is supported by the preliminary data for $pp$ collisions at 
$\sqrt{s}=7$ TeV from 
ALICE \cite{ALICE_jet_UE} indicating
that the jet FFs become softer with increase of the UE multiplicity.

In \cite{Z_pp,Z_RPP} we have addressed jet quenching in $pp$ collisions 
for light hadrons within the model of jet quenching of \cite{RAA08}
(see also \cite{RAA11,RAA12,RAA13})
based on the light-cone path integral (LCPI) approach \cite{LCPI} 
to induced gluon emission. 
We studied the medium modification of 
the $\gamma$-triggered and inclusive FFs in $pp$ collisions \cite{Z_pp}
and evaluated $R_{pp}$ for charged hadrons \cite{Z_RPP}.
It was found that the medium effects in the mini-QGP
in $pp$ collisions may be quite strong,
say, at $p_{T}\sim 10$ GeV the spectra are suppressed by
$\sim 20-30$\% at RHIC energy $\sqrt{s}=0.2$ TeV
and by $\sim 25-35$\% at LHC energy $\sqrt{s}=2.76$ TeV.

In the present letter we study flavor dependence of the $R_{pp}$.
It is of great interest for understanding the accuracy of the  pQCD 
predictions 
for heavy quark production in $pp$
collisions. Also, the flavor dependence of $R_{pp}$ is important
for theoretical predictions of $R_{AA}$ for heavy flavors. 
The question of jet quenching for heavy flavors 
has attracted much attention in recent years 
(see \cite{Renk_HQ} for a short review) 
due to the observation of strong suppression  
of single (non-photonic) electrons from decays of the heavy mesons
in experiments at RHIC \cite{PHENIX1_e,STAR_e} 
(the ``heavy quark puzzle'').
The recent measurements at LHC of $R_{AA}$ for 
electrons \cite{ALICE_e} and for $D$ mesons
\cite{ALICE_RAA_D1}
also show a strong suppression effect.
It seems to be difficult to reconcile with the
expected dead cone suppression  
of the radiative energy loss for heavy quarks predicted in 
\cite{DK}.
%%%%%%%%%%%%%%%%%%%%%%%%%%%%%%%%%%%%%%%%%%%%%%%%%%%%%
A subsequent reanalysis \cite{AZ} of the quark mass dependence
of induced gluon radiation within the LCPI approach \cite{LCPI}
demonstrated that due to the quantum finite-size effects,
ignored in the dead cone model \cite{DK}, 
the quark mass suppression of radiative energy loss 
at low energies ($\lsim 20-30$ GeV) turns out to be significantly weaker than 
predicted in \cite{DK}. And at energies $\gsim 100$ GeV
the radiative energy loss even rises with the quark mass.
Calculations of $R_{AA}$ for electrons and $D$ mesons in the scenario
without the QGP formation in $pp$ collisions performed in \cite{RAA12,RAA13}
using the LCPI approach \cite{LCPI} 
have shown reasonable agreement with experimental
data. 
However,  the experimental error bars 
for $R_{AA}$ for heavy flavors are very large. Also, the experimental 
data  (especially for the non-photonic electrons from RHIC 
\cite{PHENIX1_e,STAR_e}) are restricted
to relatively low  $p_T$, where the assumption
of dominance of the radiative energy loss and the relativistic approximation
$m_{Q}/E_Q\ll 1$ for heavy quarks (especially for $b$-quark) 
used in \cite{RAA12,RAA13} 
may be invalid. Therefore, it is probably too early to conclude that 
the ``heavy quark puzzle'' is solved, and the flavor dependence of 
the nuclear modification factors  deserves further investigation.   
In the present work we study possible effect of the mini-QGP 
formation in $pp$ collisions 
on the heavy-to-light ratios of the nuclear
modification factors in $AA$ collisions, 
which is connected to reduction of the medium suppression
for heavy quarks in $pp$ collisions. 

\section{Sketch of the theoretical framework}
We simulate jet quenching 
in $pp$ collisions within the approach
of \cite{Z_RPP}. It is qualitatively similar to the scheme 
developed for $AA$ collisions in \cite{RAA08}, which has been successfully
used in our several previous analyses of jet quenching in $AA$ collisions
\cite{RAA11,RAA12,RAA13}.
So only a brief outline of important points 
of our theoretical framework will be given here.
%So only brief highlights of our theoretical framework will be given 
%in this letter. 
The interested reader is directed to 
\cite{RAA08,Z_RPP} for details.

For the perturbative inclusive $pp$-cross section 
we use the standard formula 
\bea
\frac{d\sigma_{pert}(pp\to hX)
}{d\pb_{T} dy}=
\sum_{i}\int_{0}^{1} \frac{dz}{z^{2}}\nonumber
\\
\times
D_{h/i}^{}(z, Q)
\frac{d\sigma(pp\to iX)}{d\pb_{T}^{i} dy}\,,
\label{eq:20}
\eea
where $\pb_{T}$ is the particle transverse momentum, $y$ is rapidity 
(we will consider the central region $y=0$),
$D_{h/i}$ is the vacuum parton$\to$particle  FF, and 
${d\sigma(pp\to iX)}/{d\pb_{T}^{i} dy}$ 
is the ordinary LO pQCD hard cross section,
$\pb_{T}^{i}=\pb_{T}/z$ is the initial parton 
transverse momentum. For the initial virtuality we take $Q=p^{i}_{T}$. 
In calculating $R_{pp}$ we write the real medium modified
$pp$ cross section, that enters the numerator of (\ref{eq:10}), in
a form similar to (\ref{eq:20})
\bea
\frac{d\sigma_{m}(pp\rightarrow h X)}{d\pb_{T} dy}=
\sum_{i}\int_{0}^{1} \frac{dz}{z^{2}}\nonumber\\
\times
{D}_{h/i}^{m}(z, Q)
\frac{d\sigma(pp\rightarrow i X)}{d\pb_{T}^{i} dy}\,,\,\,\,
\label{eq:30}
\eea
where ${D}_{h/i}^{m}$ is now the medium-modified FF 
for $i\to h$ transition in the presence of the mini-QGP fireball.
It is implicit that ${D}_{h/i}^{m}$ is averaged over the jet 
production point, and the impact parameter of  $pp$ collision.
As in \cite{Z_RPP}, we used the distribution of hard processes in 
the impact parameter plane obtained using for 
the parton distribution in the transverse plane the transverse 
density distribution for quarks in the MIT bag model.

In the standard scenario without the QGP formation
the nuclear modification factor for $AA$ collisions
(we denote it $R_{AA}^{st}$) in the scheme of \cite{RAA08} 
for a given impact parameter $b$  reads
\beq
R_{AA}^{st}(b,\pb_T,y)=\frac{{dN(A A\rightarrow hX)}/{d\pb_{T}dy}}
{T_{AA}(b){d\sigma_{pert}(NN\rightarrow hX)}/{d\pb_{T}dy}}\,,
\label{eq:40}
\eeq
where 
$T_{AA}(b)=\int d\ro T_{A}(\ro) T_{A}(\ro-\bb)$,
$T_{A}$ is the nucleus profile function, and
the numerator of (\ref{eq:40})
(we omit the argument $b$) is given by
\bea
\frac{dN(AA\rightarrow hX)}{d\pb_{T} dy}=\int d\ro T_{A}(\ro)T_{A}(\ro-\bb)
\nonumber\\
\times
\frac{d\sigma_{m}(NN\rightarrow hX)}{d\pb_{T} dy}\,.
\label{eq:50}
\eea
Here  
${d\sigma_{m}(NN\to hX)}/{d\pb_{T} dy}$ is the medium-modified 
$NN\to hX$ cross section, which is given by   
the formula (\ref{eq:30}) for $NN$ collisions with the medium-modified FF for 
the QGP produced in $AA$ collision. 
In the scenario with the mini-QGP formation in $NN$ collisions 
the theoretical $R_{AA}$  (which should be compared with
the experimental $R_{AA}$) can be written as
\beq
R_{AA}=R_{AA}^{st}/R_{pp}\,.
\label{eq:60}
\eeq
Because experimentally $R_{AA}$ is defined as the ratio of 
the measured $AA\to hX$ cross section to the binary scaled 
experimental $NN\to hX$ cross section, and the latter includes the 
FSI effects in the mini-QGP produced in $NN$ collisions
(hereafter
we ignore the difference between the FSI effects in $pp$, $pn$ and $nn$ 
collisions). 

As in \cite{RAA08},  
we evaluate ${D}_{h/i}^{m}$ for each fast parton trajectory  as 
\beq
{D}_{h/i}^{m}(Q)= D_{h/j}(Q_{0})
\otimes D_{j/k}^{in}\otimes D_{k/i}(Q)\,,
\label{eq:70}
\eeq
where $\otimes$ denotes $z$-convolution, 
$D_{k/i}$ is the parton DGLAP $i\to k$ FF,
$D_{j/k}^{in}$ is the parton $j\to k$  FF in the QGP
accounting for parton energy loss, and 
$D_{h/j}$ is FF for the parton $j\to h$ transition outside 
the QGP. For the stage outside the QGP
for light partons we use for the $D_{h/j}(Q_{0})$ the 
KKP \cite{KKP} FFs  with $Q_{0}=2$ GeV.
And for the FFs $c\to D$ and $b\to B$ we use 
the Peterson parametrization 
%\cite{??}
\beq
D_{H/Q}(z)\propto \frac{1}{z[1-(1/z)-\epsilon_Q/(1-z)]^{2}}
\label{eq:80}
\eeq
with $\epsilon_{c}=0.06$ and $\epsilon_{b}=0.006$.
The DGLAP FFs have been obtained with the help of the PYTHIA event 
generator \cite{PYTHIA}.
The medium-modified FFs were computed using 
the induced gluon spectrum in the form obtained in 
\cite{Z04_RAA}.
We account for the (relatively small \cite{Z_Ecoll}) 
collisional energy loss by redefining the initial QGP 
temperature in our formulas for the medium-modified FFs 
related to induced gluon emission (see \cite{RAA08} for details).

For the  
quasiparticle masses of light quarks and gluon in the QGP 
we use the values $m_{q}=300$ and $m_{g}=400$ MeV  supported by 
the analysis of the lattice data \cite{LH},  
for $c$ and $b$ quark masses we take $m_{c}=1.2$ GeV
and $m_{b}=4.75$ GeV. We use 
the Debye mass $\mu_D$ in the QGP obtained in the lattice 
analysis \cite{Bielefeld_Md}, which gives $\mu_{D}/T$ slowly 
decreasing with $T$  
($\mu_{D}/T\approx 3$ at $T\sim 1.5T_{c}$, $\mu_{D}/T\approx 2.4$ at 
$T\sim 4T_{c}$).

We use (both for radiative and collisional energy loss) 
running $\alpha_{s}$ frozen at low momenta at
some value $\alpha_{s}^{fr}$. 
For gluon emission in vacuum for this parametrization a 
reasonable choice is $\alpha_{s}^{fr}\approx 0.7-0.8$
\cite{NZ_HERA,DKT}. 
Since thermal effects can suppress the in-medium QCD coupling, 
we treat $\alpha_{s}^{fr}$  as a free parameter
(which may differ for $pp$ and $AA$ collisions).
It is the only parameter that controls the strength
of the medium effects in our calculations.

As in \cite{RAA08},  
we use for the QGP evolution 1+1D Bjorken's model (both for 
$pp$ and $AA$ collisions), which
for the ideal gas model gives $T_{0}^{3}\tau_{0}=T^{3}\tau$, where $\tau_0$
is the thermalization time. We take $\tau_{0}=0.5$ fm. 
For $\tau<\tau_{0}$ we take medium density $\propto \tau$. 
For simplicity we neglect variation of $T_{0}$ with the 
transverse coordinates.

We fix the initial temperature of the plasma fireball in $pp$
collisions from the initial entropy density determined via
the experimental UE multiplicity density
\beq
s_{0}=\frac{C}{\tau_{0}\pi R_{f}^{2}}\frac{dN_{ch}}{d\eta}\,.
\label{eq:90}
\eeq
Here $C=dS/dy{\Big/}dN_{ch}/d\eta\approx 7.67$ \cite{BM-entropy} 
is the entropy/multiplicity ratio,
and $R_{f}$ is the typical radius of the created fireball.
We use for $R_{f}$ the parametrization of \cite{RPP} obtained from results
of simulations of $pp$ collisions  performed in \cite{glasma_pp} 
within the IP-Glasma model.
This procedure gives \cite{RPP}
$R_{f}[\sqrt{s}=0.2,2.76\,\, \mbox{TeV}]
\approx[1.3,1.44]\,\,\mbox{fm}$ and
$
T_{0}[\sqrt{s}=0.2,2.76\,\,\mbox{TeV}]
\approx[199,217]\,\,\mbox{MeV}$.
(we use the ideal gas model of the QGP with $N_{f}=2.5$).
Note that the initial temperatures would be higher by
$\sim 10-15$\% if we used the entropy from
the lattice calculations \cite{T_c2}.
We ignore this fact, because in our analysis the temperature 
is an auxiliary quantity characterizing the QGP entropy. 
The entropy is this quantity that is only important
from the point of view of parton energy loss (if we assume 
that the number density of the color constituents in the QGP is 
approximately proportional
to its entropy density). And we determine it directly from the experimental
UE multiplicity density.
For $AA$ collisions $T_{0}$ was fixed using the data on the charged 
hadron multiplicity pseudorapidity density $dN_{ch}/d\eta$ 
from RHIC \cite{STAR_Nch} and LHC \cite{CMS_Nch}.
It gives 
$T_{0}\approx 320$ MeV for central
Au+Au collisions at $\sqrt{s}=200$ GeV, and
$T_{0}\approx 420$ MeV for central
Pb+Pb collisions at $\sqrt{s}=2.76$ TeV.

Note that our results for $R_{pp}$ are not very sensitive to 
the size of the fireball, which is not well determined. 
An analysis of the stability of $R_{pp}$ 
to variation of $R_{f}$ performed in \cite{Z_RPP} has shown that 
$\pm 30$\% change in $R_{f}$ gives a small effect on $R_{pp}$.
This is due to a compensation 
between the rise of the energy loss with the fireball size 
(for a fixed density)
and its suppression with decrease of the fireball density.
In \cite{Z_RPP} we argued that the stability of $R_{pp}$ 
to variation of $R_{f}$ allows to expect that
the transveses expansion of the medium and  
the variation of the initial medium density 
in the transverse coordinates  should not have a significant 
impact on our results for $R_{pp}$.

\section{Numerical results}
In Fig.~1 we present the results of our calculations of $R_{pp}$ 
for pions, $D$ and $B$ mesons for RHIC energy  $\sqrt{s}=0.2$ TeV 
and LHC energy $\sqrt{s}=2.76$ TeV
obtained for $\alpha_{s}^{fr}=0.5$, $0.6$ and $0.7$.
For pions and $D$ mesons we show the results for $p_T>5$ GeV, and for $B$ 
mesons for the lower limit we take $p_T=7$ GeV to avoid possible problems with
applicability of the relativstic approximation and with treating the 
collisional energy loss as a small perturbation to   
the radiative energy loss for $b$ quark \cite{RAA08}.
One can see that the suppression effect due to
the mini-QGP formation turns out to be quite large. 
At $p_T\sim 10$ GeV the pion spectrum 
is suppressed by $\sim 20-30\,(25-35)$\% for RHIC (LHC), for $D$ mesons
the suppression is smaller by a factor $0.7-0.8$, and for $B$ mesons
the effect is smaller by a factor of $\sim 0.5$.
At RHIC energy all the $R_{pp}$ flatten at $p_T\gsim 15-20$ GeV.
This occures due to compensation between the effects from the reduction 
with parton energy 
of the relative parton energy loss $\Delta E/E$ (which increases
$R_{pp}$) and from the increase of
the effective power $n_{eff}$  in the power low dependence
of the hard cross section $d\sigma/dp_T\propto p_T^{-n_{eff}}$ 
(which reduces $R_{pp}$) with increase of parton energy.
Fig.~1 shows that the medium suppression in $pp$ collisions is not 
very sensitive to $\alpha_s^{fr}$. It is a consequence of 
relatively large transverse momenta $k_T$ (relative to the initial
fast parton momentum) of the emitted gluons
for the small-size QGP, because they approximately scale with the QGP size, 
$L_{QGP}$, as $1/L_{QGP}$ \cite{Z_RPP}.
In \cite{Z_RPP} in calculating the nuclear modification factor $R_{AA}$ 
for the scenario with mini-QGP in $pp$ collisions we
used the value $\alpha_s^{fr}=0.6$ in calculation of the factor $R_{pp}$.
In the present analysis we also use this value.

\begin{figure}[ht]
\epsfig{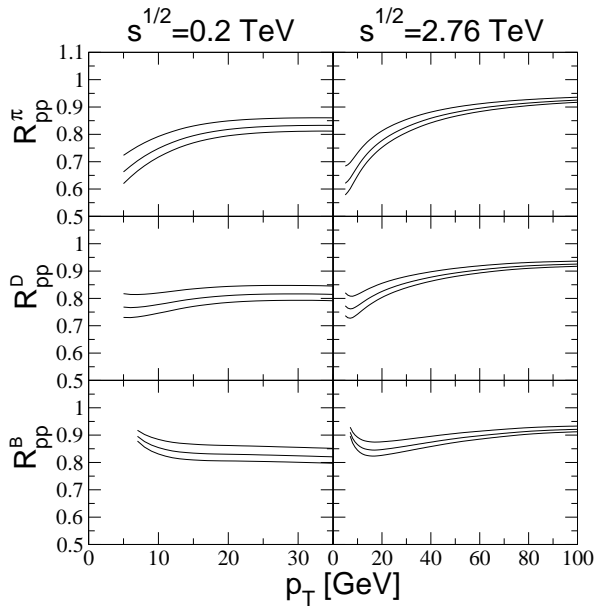}
\caption{\small $R_{pp}$ of pions, $D$ and $B$ mesons  
at $\sqrt{s}=0.2$ TeV (left panels) and $\sqrt{s}=2.76$ TeV (right panels).
From top to bottom, the curves corresponds to
$\alpha_{s}^{fr}=0.5$, $0.6$ and $0.7$.
}
\end{figure}

Even though the medium suppression of the spectra from the QGP 
formation in $pp$ collisions may be sizeable,  
it is hardly possible to observe it directly by comparing 
the experimental spectra with the theretical predictions, because
the uncertainties of the standard pQCD predictions  
(see for example\cite{Aurenche_NLO,HQ_RHIC,HQ_LHC})
are larger than the expected medium effect.
For this reason it is difficult to disentangle the medium suppression 
and other effects that can change the pQCD predictions (such as modification
of the PDFs and FFs, change of the renormalization and factorization scales)
However, as was mentioned in Introduction,
the preliminary ALICE data \cite{ALICE_jet_UE} on the UE 
multiplicity dependence of the jet FFs for $pp$ collisions at 
$\sqrt{s}=7$ TeV support the jet quenching in $pp$ 
collisions.

From Fig.~1 one sees that the flavor hierarchy 
$R_{pp}^\pi<R_{pp}^D<R_{pp}^B$
is held for $p_T\lsim 20$ GeV for RHIC and for $p_T\lsim 70$ GeV
for LHC. One can expect that in these $p_T$-regions 
the presence of the $R_{pp}$ in the formula
(\ref{eq:60}) for the nuclear modification factors in $AA$ collisions
will reduce the ratio of the nuclear modification factors
$R_{AA}$ for heavy and light hadrons.
However, to reach a definite conclusion about the effect of $R_{pp}$
on the heavy-to-light ratios of the nuclear modification factors $R_{AA}$ 
for $AA$ collisions 
one should account for the fact that in the scenario with mini-QGP
formation in $pp$ collisions the value of $\alpha_s^{fr}$
should be somewhat increased to keep stable $R_{AA}$ for pions
(since it is assumed to be adjusted to the experimental $R_{AA}$ of pions).

Comparison with the data on $R_{AA}$ for light hadrons
carried out in \cite{Z_RPP} has shown 
that for the standard scenario the RHIC data on $R_{AA}$ for pions in
Au+Au collisions for $0-5$\% centrality bin at $\sqrt{s}=0.2$ TeV  
support  $\alpha_{s}^{fr}\approx 0.5$,
and the LHC data on $R_{AA}$ for charged hadrons in Pb+Pb
collisions for $0-5$\% centrality bin  at $\sqrt{s}=2.76$ TeV support $\alpha_{s}^{fr}\approx 0.4$.
And for the scenario
with mini-QGP formation in $pp$ collisions the experimental data prefer
somewhat larger values: $\alpha_{s}^{fr}\approx 0.6\,(0.5)$ for RHIC (LHC).
For these values of $\alpha_s^{fr}$  at $p_T\gsim 10$ GeV 
the  decrease of $R_{AA}^{st}$ in the numerator of (\ref{eq:60}) 
reasonably compensates the presence of $R_{pp}<1$ in the denominator
of (\ref{eq:60}). This allows one to have approximately
the same $R_{AA}$ in both scenarios of $pp$ collisions. 
However, this
procedure acts in a different way for heavy quarks. It results
in some differences in the heavy-to-light ratios of the nuclear modification
factors $R_{AA}$ for the scenarios with and without mini-QGP production 
in $pp$ collisions.  
\begin{figure}%[ht]
\epsfig{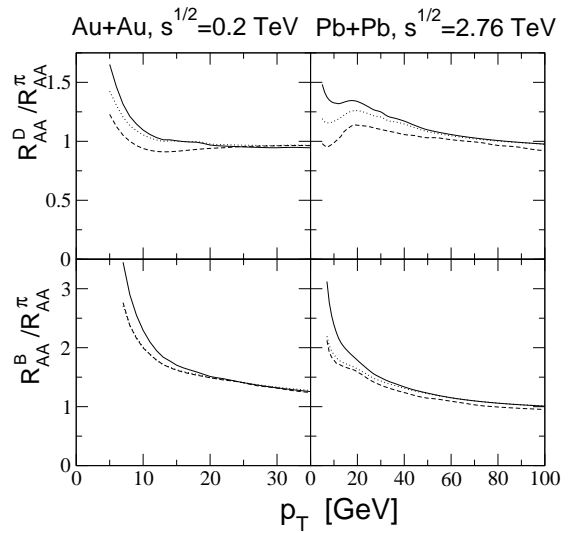}
\caption{\small
The ratios $R_{AA}^{D}/R_{AA}^{\pi}$ (upper panels)
and $R_{AA}^{B}/R_{AA}^{\pi}$ (lower panels) for $0-5$\% central
Au+Au  collisions at $\sqrt{s}=0.2$ TeV (left)
and for $0-5$\% central  Pb+Pb collisions at $\sqrt{s}=2.76$ TeV (right).
The solid curves are for $R_{AA}$ for $\alpha_{s}^{fr}=0.5$ (left) and 
$\alpha_{s}^{fr}=0.4$ (right) without $R_{pp}$ factors in (\ref{eq:60}). 
The dashed curves are for $R_{AA}$ for $\alpha_{s}^{fr}=0.6$ (left) 
and $\alpha_{s}^{fr}=0.5$ (right) with $R_{pp}$ factors in (\ref{eq:60}).
The dotted curves are for $R_{AA}$ for $\alpha_{s}^{fr}=0.5$ (left) 
and $\alpha_{s}^{fr}=0.4$ (right) with $R_{pp}$ factors in (\ref{eq:60}).
For the dashed and dotted curves the factors 
$R_{pp}$ for light and heavy flavors are calculated with $\alpha_{s}^{fr}=0.6$.}
\end{figure}

In Fig. 2  we show the ratios $R_{AA}^D/R_{AA}^\pi$ and
$R_{AA}^B/R_{AA}^\pi$ for Au+Au collisions at $\sqrt{s}=0.2$ TeV
and Pb+Pb collisions at $\sqrt{s}=2.76$ TeV  
(for $0-5$\% centrality class) obtained for $R_{AA}$ without (solid curves) 
and with (dashed curves) inclusion
of the $R_{pp}$ factors for $0-5$\% central $AA$ collisions
at RHIC and LHC. We have used the above values of $\alpha_s^{fr}$:
$0.5\,(0.4)$ for RHIC (LHC) in the standard scenario, and 
$0.6\,(0.5)$ for RHIC (LHC) in the scenario with the mini-QGP formation
in $pp$ collisions. 
From Fig.~2 one sees that the mini-QGP formation can sizeably
reduce the heavy-to-light ratios at $p_T\lsim 20$ 
GeV for RHIC and at $p_T\lsim 50$ 
GeV for LHC.
At $p_T\sim 10$ GeV (interseting in the context of the present 
situation with the ``heavy quark puzzle'')
in the mini-QGP scenario the
ratio $R_{AA}^D/R_{AA}^\pi$ becomes smaller by $\sim 20-30$\%,
and $R_{AA}^B/R_{AA}^\pi$ by $\sim 15-25$\%.
To illustrate better the role of the increase
of $\alpha_s^{fr}$ in the mini-QGP scenario, we show in  Fig.~2 
the ratios for $R_{AA}$ obtained with inclusion of $R_{pp}$ 
in (\ref{eq:60}), but for $R_{AA}^{st}$ evaluated 
without renormaization of $\alpha_s^{fr}$ (dotted curves).
By comparing the results for two choices of $\alpha_s^{fr}$  
one can see that the renormalization of
$\alpha_s^{fr}$ gives a sizeable reduction of $R_{AA}^D/R_{AA}^\pi$, but
practically does not change $R_{AA}^B/R_{AA}^\pi$. 
Thus, from Fig.~2 we see that the mini-QGP formation in $pp$
collisions weakens the flavor dependence of the nuclear
modification factors $R_{AA}$ at RHIC and LHC energies.
However, the effect is not very strong, and due to rather large
experimental error bars for the available heavy flavor $R_{AA}$ data 
\cite{PHENIX1_e,STAR_e,ALICE_e,ALICE_RAA_D1}
the situation with their description
in the scenario with the mini-QGP production in $pp$ collisions does not differ
significantly from our previous analysis \cite{RAA12,RAA13} within the standard
scenario without the QGP formation in $pp$ collisions.
For this reason we do not present a comparison
with experiment in the present letter. Anyway, for a conclusive
comparison between theory and experiment we need more accurate data
on the $R_{AA}$ for heavy flavors at higher $p_T$.

\section{Summary}
The medium FSI effects should modify high-$p_T$ jets in $pp$ 
collisions if the mini-QGP production occurs.  
We have studied the flavor dependence of the medium modification
factor $R_{pp}$ in this scenario. We evaluated $R_{pp}$ for pions,  
$D$ and $B$ mesons for RHIC ($\sqrt{s}=0.2$ TeV) and LHC ($\sqrt{s}=2.76$ TeV) 
energies. We have observed that at $p_T\sim 10$ GeV
the pion spectrum  is suppressed by $\sim 20-30\,(25-35)$\% for RHIC (LHC), 
for $D$~($B$) mesons the effect is smaller by a factor of $\sim 0.7-0.8$ 
($0.5$) than for pions.
The flavor hierarchy 
$R_{pp}^\pi<R_{pp}^D<R_{pp}^B$
is held at $p_T\lsim 20$ GeV for RHIC and at $p_T\lsim 70$ GeV
for LHC. We demonstrated that this gives a sizeable reduction of the 
heavy-to-light ratios of the nuclear modifcation factors $R_{AA}$.
At $p_T\sim 10$ GeV the ratio $R_{AA}^{D}/R_{AA}^{\pi}$ is suppressed
by $\sim 20-30$\% and $R_{AA}^B/R_{AA}^\pi$ by $\sim 15-25$\%
as compared to that in the standard scenario without
the QGP production in $pp$ collisions.

\begin{acknowledgments}
This work is supported 
in part by the 
grant RFBR
15-02-00668-a
and the program SS-3139.2014.2.
\end{acknowledgments}


\section*{References}
\begin{thebibliography}{99}

\bibitem{Z_pp}
%Medium modification of photon-tagged and inclusive 
%jets in high-multiplicity proton-proton collisions
B.G. Zakharov, Phys.~Rev.~Lett. {\bf 112}, 032301 (2014) 
[arXiv:1307.3674].




\bibitem{Bozek_pp} 	
%Observation of the collective flow in proton-proton collisions
P.~Bozek,
Acta Phys. Polon. B{\bf 41}, 837 (2010)
[arXiv:0911.2392].
%Detailed record - Cited by 25 records

\bibitem{Wied_pp}	
%Eccentricity fluctuations make flow measurable in high 
%multiplicity p-p collisions
J.~Casalderrey-Solana and U.A.~Wiedemann,
Phys. Rev. Lett. {\bf 104}, 102301 (2010) 
[arXiv:0911.4400].


\bibitem{Camp1} 	
%Experimental equation of state in proton-proton and proton-antiproton 
%collisions and phase transition to quark gluon plasma
R.~Campanini, G.~Ferri, and G.~Ferri,
Phys. Lett. B{\bf 703}, 237 (2011)
[arXiv:1106.2008].


\bibitem{Gyulassy_pp}
%Can hyperon/meson ratios in rare high multiplicity pp 
%collisions at Large Hadron Collider energies provide 
%signatures of mini-quark-gluon plasma formation?
V.~Topor Pop, M.~Gyulassy, J.~Barrette, C.~Gale, and A.~Warburton,
Phys.~Rev. C{\bf 86}, 044902  (2012) 
[arXiv:1203.6679].

\bibitem{glasma_pp}
%Initial state geometry and the role of hydrodynamics in proton-proton, 
%proton-nucleus and deuteron-nucleus collisions
A.~Bzdak, B.~Schenke, P.~Tribedy, and R.~Venugopalan,	
Phys.~Rev. C{\bf 87}, 064906 (2013)
[arXiv:1304.3403].

\bibitem{SZ}
%High Multiplicity pp and pA Collisions: 
%Hydrodynamics at its Edge and Stringy Black Hole
E.~Shuryak and I.~Zahed, 
Phys.~Rev. C{\bf 88}, 044915 (2013)
[arXiv:1301.4470].


\bibitem{CMS_ridge1}
%Observation of Long-Range Near-Side Angular Correlations 
%in Proton-Proton Collisions at the LHC
V.~Khachatryan {\sl et al.}
[CMS Collaboration],
JHEP {\bf 1009}, 091 (2010). % [arXiv:1009.4122].

\bibitem{Field}
%Min-Bias and the Underlying Event at the LHC
R.~Field,
Acta Phys.~Polon. B{\bf 42}, 2631  (2011) 
%2631-2656
[arXiv:1110.5530].


\bibitem{ALICE_jet_UE} 	
%Connecting the underlying event with jet properties in pp collisions at 
%s√ = 7 TeV with the ALICE experiment
H.L.~Vargas,  for the 
ALICE Collaboration,
J. Phys. Conf. Ser. {\bf 389}, 012004  (2012) 
%Conference: C12-04-07 Proceedings
[arXiv:1208.0940].

\bibitem{Z_RPP}
%Parton energy loss in the mini quark-gluon plasma and 
%jet quenching in proton-proton collisions
B.G. Zakharov,
J.~Phys. G{\bf 41}, 075008  (2014) 
[arXiv:1311.1159].

\bibitem{RAA08}
% Jet quenching with running coupling including radiative and 
%collisional energy losses.
B.G.~Zakharov, JETP Lett. {\bf 88}, 781 (2008). %[arXiv:0811.0445].


%Variation of jet quenching from RHIC to LHC and thermal suppression of 
%QCD coupling constant.
\bibitem{RAA11}
B.G.~Zakharov,
JETP Lett. {\bf 93}, 683 (2011) [arXiv:1105.2028].


\bibitem{RAA12}
%Nuclear modification factor for light and heavy 
%flavors within pQCD and recent data from the LHC
B.G.~Zakharov, 
JETP Lett. {\bf 96}, 616 (2013) [arXiv:1210.4148].

\bibitem{RAA13}
%Nuclear suppression of light hadrons and single electrons at RHIC and LHC
B.G. Zakharov,
J. Phys. G{\bf 40}, 085003  (2013) [arXiv:1304.5742].

\bibitem{LCPI}
B.G.~Zakharov, JETP\ Lett. {\bf 63}, 952 (1996); {\em ibid}
{\bf 65}, 615 (1997);
{\bf 70}, 176 (1999);
Phys.\ Atom.\ Nucl. {\bf 61}, 838 (1998).


\bibitem{Renk_HQ}
%Jet quenching and heavy quarks
T. Renk,
J.~Phys.~Conf.~Ser. {\bf 509}, 012022 (2014)
[arXiv:1309.3059]. 



\bibitem{PHENIX1_e}
%Nuclear modification of electron spectra and implications 
%for heavy quark energy loss in Au+Au collisions at s(NN)**(1/2) - 200-GeV.
S.S. Adler {\it et al.} [PHENIX Collaboration],
Phys. Rev. Lett. {\bf 96}, 032301 (2006).


\bibitem{STAR_e}
%Erratum: Transverse momentum and centrality dependence of high-\pt\ 
%non-photonic electron suppression in Au+Au collisions at \sqrtsNN\ = 200 GeV.
B.I. Abelev {\it et al.} [STAR Collaboration],
Phys. Rev. Lett. {\bf 98},  192301 (2007)
[arXiv:nucl-ex/0607012], Erratum-ibid. 106 (2011) 159902.



\bibitem{ALICE_e}
S. Sakai, for the ALICE Collaboration, contribution to the Quark Matter
2012 Conf., http://qm2012.bnl.gov/default.asp.

\bibitem{ALICE_RAA_D1}
%Suppression of high transverse momentum D mesons in central Pb-Pb 
%collisions at sNN‾‾‾‾√=2.76 TeV.
B. Abelev {\it et al.}
[ALICE Collaboration],
JHEP {\bf 1209}, 112 (2012) [arXiv:1203.2160].


\bibitem{DK}
Y.L.~Dokshitzer and D.E.~Kharzeev,
Phys.\ Lett B{\bf 519}, 199 (2001).

%Anomalous mass dependence of radiative quark energy 
%loss in a finite-size quark-gluon plasma.
\bibitem{AZ}
P. Aurenche and B.G. Zakharov, 
JETP Lett. {\bf 90}, 237 (2009)
[arXiv:0907.1918].


\bibitem{KKP}
B.A.~Kniehl, G.~Kramer, and B.~Potter, 
Nucl.\ Phys. B{\bf 582}, 514 (2000).

\bibitem{PYTHIA}
% Pythia 6.3 physics and manual.
T.~Sjostrand, L.~Lonnblad, S.~Mrenna, and  P.~Skands,
arXiv:hep-ph/0308153.


%Radiative parton energy loss and jet quenching in high-energy 
%heavy-ion collisions.
\bibitem{Z04_RAA}
B.G.~Zakharov, JETP Lett. {\bf 80}, 617 (2004)
[hep-ph/0410321].

\bibitem{Z_Ecoll}
%Parton energy loss in an expanding quark-gluon plasma: Radiative versus collisional.
B.G.~Zakharov,
JETP Lett. {\bf 86}, 444 (2007)
[arXiv:0708.0816].

\bibitem{LH}
P.~L\'evai and U.~Heinz,
Phys.\ Rev.\ C{\bf 57}, 1879 (1998).

% Static quark anti-quark interactions in zero and finite 
%temperature QCD. I. Heavy quark free energies, 
%running coupling and quarkonium binding.
\bibitem{Bielefeld_Md}
O.~Kaczmarek and F.~Zantow,
Phys. Rev. D{\bf 71}, 114510 (2005).





\bibitem{NZ_HERA}
%DEEP INELASTIC SCATTERING AT HERA AND THE BFKL POMERON.
N.N.~Nikolaev and B.G.~Zakharov,
Phys. Lett. B{\bf 327}, 149 (1994). 

\bibitem{DKT}
Yu.L.~Dokshitzer, V.A.~Khoze, and S.I.~Troyan,
Phys.\ Rev. D{\bf 53}, 89 (1996).

\bibitem{BM-entropy}
% From entropy and jet quenching to deconfinement?
B.~M\"uller and K.~Rajagopal,
Eur. Phys. J. C{\bf 43}, 15 (2005) [arXiv:hep-ph/0502174].



\bibitem{RPP} 	
%Transverse Momentum of Protons, Pions and Kaons in High 
%Multiplicity pp and pA Collisions: Evidence for the Color Glass Condensate?
L. McLerran, M.~Praszalowicz, and B.~Schenke,
arXiv:1306.2350.


\bibitem{T_c2}
%Thermodynamics of the QCD transition from lattice
S.~Borsanyi,
Nucl.~Phys.A{\bf 904-905}, 270c (2013) 
%270c-277c
[arXiv:1210.6901].


\bibitem{STAR_Nch}
%Systematic Measurements of Identified Particle Spectra in 
%pp,d+ Au and Au+Au Collisions from STAR
B.I. Abelev {\it et al.}
[STAR Collaboration ],
Phys. Rev. C{\bf 79}, 034909 (2009).
%e-Print: arXiv:0808.2041 [nucl-ex] | PDF 

\bibitem{CMS_Nch}
%Dependence on pseudorapidity and centrality of charged 
%hadron production in PbPb collisions at a nucleon-nucleon 
%centre-of-mass energy of 2.76 TeV.
S. Chatrchyan {\it et al.} [CMS Collaboration], 
JHEP {\bf 1108}, 141 (2011).
%[arXiv:1107.4800].




\bibitem{Aurenche_NLO} 	
%Large p(T) inclusive pi0 cross-sections and next-to-leading-order 
%QCD predictions
P.~Aurenche, M.~Fontannaz, J.P.~Guillet, B.A.~Kniehl, and M.~Werlen,
Eur.~Phys.~J. C{\bf 13}, 347 (2000) [hep-ph/9910252].


\bibitem{HQ_RHIC}
%QCD predictions for charm and bottom production at RHIC
M. Cacciari, P. Nason, and R. Vogt,
Phys.~Rev.~Lett. {\bf 95}, 122001 (2005)
%DOI: 10.1103/PhysRevLett.95.122001
[hep-ph/0502203]

\bibitem{HQ_LHC}
%Theoretical predictions for charm and bottom production at the LHC
M. Cacciari {\sl et al.}, 
%S. Frixione, N. Houdeau, M.L. Mangano, P. Nason, Giovanni Ridolfi
JHEP {\bf 1210}, 137 (2012) [arXiv:1205.6344].


\end{thebibliography}
\end{document}